\documentclass[preprint,3p]{elsarticle}
\usepackage{dsfont,amsmath}
\biboptions{sort&compress}
\journal{Physics Letters B}
\usepackage[dvips]{color}

\begin{document}
\begin{frontmatter}
\title{A Higher Order GUP with Minimal Length Uncertainty and Maximal Momentum}
\author{Pouria Pedram}
\ead{p.pedram@srbiau.ac.ir}
\address{Department of Physics, Science and Research Branch, Islamic Azad University, Tehran, Iran}

\begin{abstract}
We present a higher order generalized (gravitational) uncertainty
principle (GUP) in the form $[X,P]=i\hbar/(1-\beta P^2)$. This form
of GUP is consistent with various proposals of quantum gravity such
as string theory, loop quantum gravity, doubly special relativity,
and predicts both a minimal length uncertainty and a maximal
observable momentum. We show that the presence of the maximal
momentum results in an upper bound on the energy spectrum of the
momentum eigenstates and the harmonic oscillator.
\end{abstract}


\end{frontmatter}

\section{Introduction}
In recent years, there is a great interest to study the effects of
the Generalized Uncertainty Principle (GUP) and the Modified
Dispersion Relation (MDR) on various quantum mechanical systems (see
\cite{felder} and the references therein). Indeed, the ideas of GUP
and MDR arise naturally from various candidates of quantum gravity
such as string theory \cite{1,2,3,4}, loop quantum gravity \cite{5},
noncommutative spacetime \cite{6,7,8}, and black holes gedanken
experiments \cite{9,10}. These theories indicate that the Heisenberg
uncertainty principle should be modified to incorporate additional
constraints in the presence of the gravitational field.

The existence of a minimal length scale of the order of the Planck
length $\ell_{Pl}=\sqrt{\frac{G\hbar}{c^3}}\approx 10^{-35}m$ is one
of the main outcomes of various GUP proposals where $G$ is Newton's
gravitational constant. In fact, beyond the Planck energy scale, the
effects of gravity are so important which would result in
discreteness of the very spacetime. Notably, the quantum field
theory in curved background can be renormalizable by introducing a
minimal observable length as an effective cutoff in the ultraviolet
domain. Also, in the string-theoretic argument, we can say that the
string cannot probe distances smaller than its own length.

The introduction of this idea has drawn much attention in the
literature to study the effects of GUP on small scale and large
scale systems
\cite{14,18,19,101,banerjee,p2,p3,p4,p5,p6,NP,102,103,104,105,106,main,Nouicer,pedramPRD,pedramPLB}.
It is also possible to incorporate the idea of a maximal observable
momentum into this scenario. In fact, in doubly special relativity
(DSR) theories, we consider the Planck energy (Planck Momentum) as
an additional invariant other than the velocity of light
\cite{21,22,23}. Recently, the construction of a perturbative GUP
which is consistent with DSR theories is also discussed in
Refs.~\cite{pedram,pedram1,main1,main2,main3,KE}. It is also shown
that a minimum uncertainty in momentum can arise from curvature, as
part of a study that indicated that curvature and noncommutativity
can be seen as dual to each other \cite{Kempf1}.

In this Letter, we investigate the effects of a new generalized
uncertainty principle to all orders in the Planck length on some
quantum mechanical systems. This form of GUP implies the existence
of a minimal length uncertainty and a maximal momentum in agreement
with various theories of quantum gravity. Here, we study the
problems of the eigenstates of the position operator, maximal
localized states, and the harmonic oscillator in this framework and
obtain their energy spectrum, which as we shall see, are bounded
from above.

\section{The Generalized Uncertainty Principle}\label{sec2}
First let us consider a generalized uncertainty principle proposed
by Kempf, Mangano and Mann (KMM) and results in a minimum observable
length
\begin{eqnarray}\label{gup}
\Delta X \Delta P \geq \frac{\hbar}{2} \left( 1 +\beta (\Delta P)^2
+\zeta \right),
\end{eqnarray}
where $\beta$ is the GUP parameter and $\zeta$ is a positive
constant that depends on the expectation values of the momentum
operator, i.e., $\zeta=\beta\langle P\rangle^2$. We also have
$\beta=\beta_0/(M_{Pl} c)^2$ where $M_{Pl}$ is the Planck mass and
$\beta_0$ is of the order of the unity. It is straightforward to
check that the inequality relation (\ref{gup}) implies the existence
of a minimum observable length as $(\Delta
X)^{\mathrm{KMM}}_{min}=\hbar\sqrt{\beta}$. In one-dimension, the
above uncertainty relation can be obtained from the following
deformed commutation relation:
\begin{eqnarray}\label{gupc}
[X,P]=i\hbar(1+\beta P^2).
\end{eqnarray}
As KMM have indicated in their seminal paper, we can write $X$ and
$P$ in momentum space representation as \cite{7}
\begin{eqnarray}\label{k1}
P \phi(p)&=& p\,\phi(p),\\ X\phi(p)&=& i\hbar \left( 1 + \beta
p^2\right)\partial_p\phi(p),\label{k2}
\end{eqnarray}
where $X$ and $P$ are symmetric operators on the dense domain
$S_{\infty}$ with respect to the following scalar product:
\begin{eqnarray}\label{scalar}
\langle\psi|\phi\rangle=\int_{-\infty}^{+\infty}\frac{\mathrm{d}p}{1+\beta
p^2}\psi^{*}(p)\phi(p),
\end{eqnarray}
where $\int_{-\infty}^{+\infty}\frac{\mathrm{d}p}{1+\beta
p^2}|p\rangle\langle p|=1$ and $\langle p|p'\rangle=\left( 1 + \beta
p^2\right)\delta(p-p')$. With this definition, the commutation
relation (\ref{gupc}) is exactly satisfied.

Based on the field theory on nonanticommutative superspace, Nouicer
has suggested the following higher order GUP which agrees with
(\ref{gupc}) to the leading order and also predicts a minimal length
uncertainty
\begin{eqnarray}
[X,P]=i\hbar\exp\left(\beta P^2\right).
\end{eqnarray}
This algebra can be satisfied from the following representation of
the position and momentum operators:
\begin{eqnarray}\label{rep1-1}
P \phi(p)&=& p\,\phi(p),\\ X\phi(p)&=& i\hbar \exp\left(\beta
p^2\right)\partial_p\phi(p).\label{rep1-2}
\end{eqnarray}
Now the symmetricity condition of the position operator implies the
following modified completeness relation and scalar product
\begin{eqnarray}
\langle\psi|\phi\rangle&=&\int_{-\infty}^{+\infty}\mathrm{d}p
 \exp\left(-\beta p^2\right)\psi^{*}(p)\phi(p),\\
\langle p|p'\rangle&=& \exp\left(\beta p^2\right)\delta(p-p').
\end{eqnarray}
Also, the absolutely smallest uncertainty in position is given by
$(\Delta
X)^{\mathrm{Nouicer}}_{min}=\sqrt{\frac{e}{2}}\hbar\sqrt{\beta}$.

To incorporate the idea of the maximal momentum, Ali, Das and
Vagenas have proposed the following modified commutation relation
\cite{main1,main2,main3}
\begin{eqnarray}\label{xp}
[X_i, P_j] &=& i \hbar\bigg[ \delta_{ij}- \alpha \left( P\delta_{ij}
+ \frac{P_i P_j}{P} \right)+ \alpha^2 \left( P^2 \delta_{ij} +
3P_{i} P_{j} \right) \bigg],
\end{eqnarray}
where $\alpha = {\alpha_0}/{M_{Pl}c} = {\alpha_0 \ell_{Pl}}/{\hbar}$
is the GUP parameter, $P^{2} = \sum\limits_{j=1}^{3}P_{j}P_{j}$,
$M_{Pl}$ is the Planck mass, and $M_{Pl} c^2\sim10^{19}$GeV is the
Planck energy. This form of GUP implies both a minimal length
uncertainty and a maximal momentum uncertainty, namely \cite{main1}
\begin{eqnarray}\label{min1-1}
\Delta X &\geq& (\Delta X)_{min}  \approx \alpha_0\ell_{Pl}=\hbar \alpha,  \\
\Delta P &\leq& (\Delta P)_{max} \approx \frac{ M_{Pl}c}{
\alpha_0}=1/\alpha.\label{min1-2}
\end{eqnarray}
The commutation relation (\ref{xp}) is approximately satisfies by
the the following representation
\begin{eqnarray}\label{x0p0-1}
X_i &=& x_{i},\\
P_i &=& p_{i} \left( 1 - \alpha p + 2\alpha^2 p^2
\right),\label{x0p0-2}
\end{eqnarray}
where $x_{i}$ and $p_{i}$ obey the usual commutation relations
$[x_{i},p_{j}]=i\hbar\delta_{ij}$ and $p$ is the magnitude of
$\vec{p}$. Now Eq.~(\ref{min1-1}) implies
$\alpha\approx\sqrt{\beta}$. However, this proposal has the
following difficulties:
\begin{itemize}
  \item It is perturbative, i.e., it is only valid for small values of the GUP parameter.
  \item Although the minimal length uncertainty can be interpreted as the minimal length, the maximal momentum uncertainty
  differs from the idea of the maximal momentum which is required in DSR
  theories. Indeed Eq.~(\ref{min1-2}) puts an upper bound on the
  uncertainty of the momentum measurement, not on the value of the
  observed momentum.
  \item It does not imply noncommutative geometry, because $[X_i,X_j]=0$ [see Eq.~(\ref{x0p0-1})].
\end{itemize}

To overcome these problems, consider the following higher order
generalized uncertainty principle (GUP*) which implies both the
minimal length uncertainty and the maximal observable momentum
\begin{eqnarray}\label{guph}
[X,P]=\frac{i\hbar}{1-\beta P^2}.
\end{eqnarray}
This commutation relation agrees with KMM's and Noucier's proposals
to the leading order and contains a singularity at $P^2=1/\beta$.
This fact shows that the momentum of the particle cannot exceed
$1/\sqrt{\beta}\approx1/\alpha$ which agrees formally with
Eq.~(\ref{min1-2}). As stated before, Eqs.~(\ref{min1-2}) and
(\ref{guph}) imply two basically different quantities. However, the
presence of an upper bound on the momentum properly agrees with DSR
theories. As we shall see, the physical observables such as energy
and momentum are not only nonsingular, but also are bounded from
above.

Note that, this choice is the simplest choice (using rational
functions) that implements momentum cutoff at the commutation
relation level, and which reduces to KMM proposal. One way of
getting rational approximations from a truncated power series is by
using Pad\'{e} resummation. Indeed, the Pad\'{e} approximant is the
best approximation of a function by a rational function and for a
series expansion $f(P) = \sum^k_{i=0} f_iP^i + \cdots$ up to the
order $k$ is presented by \cite{pade}
\begin{eqnarray}
[m/n]=\frac{a_0+a_1P+\cdots+a_mP^m}{1+b_1P+\cdots+b_nP^n},\hspace{2cm}m+n=k,
\end{eqnarray}
where $a_i$ and $b_i$ are found such that the series expansion of
$[m/n]$ up to ${\mathcal O}(k)$ equals the original series, namely
\begin{eqnarray}
\sum^k_{i=0} f_iP^i=[m/n]+{\mathcal O}(m+n+1).
\end{eqnarray}
So the $m+n+1$ unknown coefficients are given uniquely by the $k +
1$ coefficients $f_i$. Now if one treats the KMM relation
(\ref{gupc}) as a low momentum $[2/0]$ approximation of the ultimate
GUP proposal $[X,P]=i\hbar f(P)$, then its $[0/2]$ Pad\'{e}
approximant gives Eq.~(\ref{guph}) which also contains an additional
property, i.e., the momentum cutoff. Of course Pad\'{e} resummations
are approximations and not a rigorous justification but they are
popular in many fields in estimating ``nonperturbative'' effects.

On the other hand, and from a physical viewpoint, GUPs are common
phenomenological aspects of all promising candidates of quantum
gravity. Adopting a mathematically well-motivated and
nonperturbative GUP has the potential to shed light on even more
phenomenological aspects of the mentioned candidates. Especially,
the relatively different algebraic structure of the GUP*
(\ref{guph}) has new implications on the Hilbert space
representation of quantum mechanics that overcomes some conceptual
problems raised in the original KMM formalism such as the divergence
of the energy spectrum of the eigenfunctions of the position
operator. Unlike the KMM case that the energy of the short
wavelength modes are divergent, it is straightforward to show that
in our case there is no divergence in the energy spectrum for short
wavelengths  [see Eq.~(\ref{posi22})]. Also, the different Hilbert
space structure may have some new implications on measurement theory
in this framework.

To satisfy the above commutation relation, we can write the position
and momentum operators in the momentum space representation as
\begin{eqnarray}\label{rep1}
P \phi(p)&=& p\,\phi(p),\\
X\phi(p)&=& \frac{i\hbar}{1 - \beta
p^2}\partial_p\phi(p).\label{rep2}
\end{eqnarray}
Using the symmetricity condition of the position operator the
modified completeness relation and scalar product can be written as
\begin{eqnarray}
\langle\psi|\phi\rangle&=&\int_{-1/\sqrt{\beta}}^{+1/\sqrt{\beta}}\mathrm{d}p
\left(1-\beta p^2\right)\psi^{*}(p)\phi(p),\\
\langle p|p'\rangle&=& \frac{\delta(p-p')}{1-\beta p^2}.
\end{eqnarray}
The uncertainty relation that arises from GUP* is given by
\begin{eqnarray}
\hspace{-1cm}(\Delta X)(\Delta
P)&\geq&\left\langle\frac{\hbar/2}{1-\beta
P^2}\right\rangle,\nonumber\\
&\geq&\frac{\hbar}{2}\left(1+\beta \left\langle
P^2\right\rangle+\beta^2 \left\langle P^4\right\rangle+\beta^3
\left\langle
P^6\right\rangle+\cdots\right),\nonumber\\
&\geq&\frac{\hbar}{2}\left(1+\beta \left\langle
P^2\right\rangle+\beta^2 \left\langle P^2\right\rangle^2+\beta^3
\left\langle P^2\right\rangle^3+\cdots\right),\nonumber\\
&\geq&\frac{\hbar}{2}\Big(1+\beta \left[ (\Delta P)^2+\langle
P\rangle^2\right]+\beta^2 \left[ (\Delta P)^2+\langle
P\rangle^2\right]^2+\beta^3 \left[ (\Delta P)^2+\langle
P\rangle^2\right]^3+\cdots\Big),\nonumber\\
&\geq&\frac{\hbar/2}{1-\beta \left[ (\Delta P)^2+\langle
P\rangle^2\right]},
\end{eqnarray}
where we have used the property $\langle P^{2n}\rangle \geq \langle
P^{2}\rangle^n $. In order to find the minimal length uncertainty of
this deformed algebra, we consider the physical states for which we
have $\langle P\rangle= 0$ and solve the following saturate GUP* for
$\Delta P$
\begin{eqnarray}
(\Delta X)(\Delta P)=\frac{\hbar/2}{1-\beta (\Delta P)^2},
\end{eqnarray}
which has a minimum at $\Delta P=1/\sqrt{3\beta}$. So the absolutely
smallest uncertainty in position is given by
\begin{eqnarray}
(\Delta X)^*_{min}=\frac{3\sqrt{3}}{4}\hbar\sqrt{\beta}.
\end{eqnarray}
In Table \ref{tab1}, we have compared minimal length uncertainties
from various GUP scenarios. These results show that $(\Delta
X)^{\mathrm{KMM}}_{min}<(\Delta X)^{\mathrm{Noucier}}_{min}<(\Delta
X)^*_{min}$.

\begin{table}
\centering
\begin{tabular}{cccc}\hline
&KMM & Nouicer & GUP* \\\hline
$(\Delta X)_{min}$& $\displaystyle\hbar\sqrt{\beta}$ &$\displaystyle\sqrt{\frac{e}{2}}\hbar\sqrt{\beta}$& $\displaystyle\frac{3\sqrt{3}}{4}\hbar\sqrt{\beta}$ \\
$P_{max}$& -- &--& $\displaystyle\frac{1}{\sqrt{\beta}}$ \\\hline
\end{tabular}
\caption{\label{tab1}The minimal length uncertainties and maximal
momentums in three GUP frameworks.}
\end{table}

\section{Functional analysis of the position operator}
\begin{figure}
\begin{center}
\includegraphics[width=7cm]{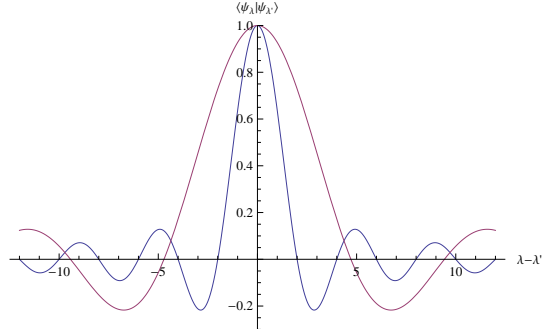}
\caption{\label{fig1}Plotting
$\langle\psi_\lambda|\psi_{\lambda'}\rangle$ over $\lambda-\lambda'$
in units of $\hbar\sqrt{\beta}$ for the KMM GUP (blue line) and GUP*
(red line).}
\end{center}
\end{figure}

The eigenvalue problem for the position operator in the GUP*
framework and in the momentum space is given by
\begin{eqnarray}
\frac{i\hbar}{1-\beta
p^2}\partial_p\psi_\lambda(p)=\lambda\psi_\lambda(p).
\end{eqnarray}
This equation can be solved to obtain the position eigenvectors
\begin{eqnarray}
\psi_\lambda(p)=c \exp\left(\frac{-i\lambda
p}{\hbar}\left(1-\frac{\beta}{3}p^2\right)\right).
\end{eqnarray}
The eigenfunctions are normalizable
\begin{eqnarray}
1=cc^*\int_{-1/\sqrt{\beta}}^{+1/\sqrt{\beta}}\mathrm{d}p
\left(1-\beta p^2\right)=\frac{4cc^*}{3\sqrt{\beta}}.
\end{eqnarray}
Therefore
\begin{eqnarray}
\psi_\lambda(p)=\frac{\sqrt{3\sqrt{\beta}}}{2}
\exp\left(\frac{-i\lambda
p}{\hbar}\left(1-\frac{\beta}{3}p^2\right)\right).
\end{eqnarray}
Now we calculate the scalar product of the position eigenstates
\begin{eqnarray}
\langle\psi_\lambda|\psi_{\lambda'}\rangle&=&\frac{3\sqrt{\beta}}{4}\int_{-1/\sqrt{\beta}}^{+1/\sqrt{\beta}}
\left(1-\beta p^2\right)\exp\left(\frac{i(\lambda-\lambda')
p}{\hbar}\left(1-\frac{\beta}{3}p^2\right)\right)\mathrm{d}p,\nonumber\\
&=&\frac{3\hbar\sqrt{\beta}}{2(\lambda-\lambda')}\sin\left(\frac{2(\lambda-\lambda')}{3\hbar\sqrt{\beta}}\right).\label{innerLambda}
\end{eqnarray}
Thus, similar to the KMM scenario, the position eigenstates are
generally no longer orthogonal. In Fig.~\ref{fig1}, we have depicted
$\langle\psi_\lambda|\psi_{\lambda'}\rangle$ for the KMM proposal and
GUP*. Although this quantity in both models has a same functional
form, it is more oscillatory in the KMM framework.

\subsection{Maximal localization states}
The maximal localization states $|\psi^{\mathrm{ML}}_\xi\rangle$ are
defined with the properties
\begin{eqnarray}\label{miangin}
\langle \psi^{\mathrm{ML}}_\xi|X|\psi^{\mathrm{ML}}_\xi\rangle=\xi,
\end{eqnarray}
and
\begin{eqnarray}\label{deltaXML}
\Delta X_{|\psi^{\mathrm{ML}}_\xi\rangle}=(\Delta X)^*_{min}.
\end{eqnarray}
These states also satisfy \cite{7}
\begin{eqnarray}
\bigg(X-\langle X\rangle+\frac{\langle[X,P]\rangle}{2(\Delta
P)^2}\left(P-\langle P\rangle\right)\bigg)|\psi\rangle=0.
\end{eqnarray}
To proceed further we need to express $\langle[X,P]\rangle$ in terms
of $\Delta P$ and $\langle P\rangle$. However, since
$\langle[X,P]\rangle$ also depends on $\langle P^4\rangle$, $\langle
P^6\rangle$, etc., and these quantities cannot be calculated before
specifying $|\psi\rangle$, to first order in the GUP parameter we
can use the approximate relation $\langle[X,P]\rangle\simeq
i\hbar\left(1+\beta (\Delta P)^2+\beta\langle P\rangle^2 \right)$.
So, in momentum space, the above equation takes the form
\begin{eqnarray}
&\bigg(&\frac{i\hbar}{1-\beta p^2}\frac{\partial}{\partial
p}-\langle X\rangle+i\hbar\frac{1+\beta (\Delta P)^2+\beta\langle
P\rangle^2}{2(\Delta P)^2}\Big(p-\langle
P\rangle\Big)\bigg)\psi(p)\simeq0,
\end{eqnarray}
which has the solution
\begin{eqnarray}
\psi(p)\simeq &&\mathcal{N}\exp\bigg[\left(-\frac{i}{\hbar}\langle
X\rangle+\frac{1+\beta (\Delta P)^2+\beta\langle
P\rangle^2}{2(\Delta P)^2}\langle P\rangle\right)\nonumber\\
&&\times\left(p-\frac{\beta}{3}p^3\right)-\frac{1+\beta (\Delta
P)^2+\beta\langle P\rangle^2}{4(\Delta P)^2}
\left(p^2-\frac{\beta}{2}p^4\right)\bigg].
\end{eqnarray}
To find the absolutely maximal localization states we need to choose
the critical momentum uncertainty $\Delta P=1/\sqrt{3\beta}$ that
gives the minimal length uncertainty and take $\langle P\rangle=0$,
i.e.,
\begin{eqnarray}\label{psiML}
\hspace{-1cm}\psi^{\mathrm{ML}}_\xi(p)\simeq
\mathcal{N}\exp\bigg[-\frac{i}{\hbar}\xi\left(p-\frac{\beta}{3}p^3\right)-\beta
\left(p^2-\frac{\beta}{2}p^4\right)\bigg],
\end{eqnarray}
where the normalization factor is given by
\begin{eqnarray}
1&=&\mathcal{N}\mathcal{N}^*\int_{-1/\sqrt{\beta}}^{+1/\sqrt{\beta}}\mathrm{d}p
\left(1-\beta p^2\right)\exp\left(2\beta p^2-\beta^2
p^4\right),\nonumber\\
&=&1.0123\frac{\mathcal{N}^2}{\sqrt{\beta}}.
\end{eqnarray}
Note that $\psi^{\mathrm{ML}}_\xi(p)$ exactly satisfies
Eq.~(\ref{miangin}). However, because of the approximation that
assumed to find $\psi^{\mathrm{ML}}_\xi(p)$ (\ref{psiML}), it
approximately obeys relation (\ref{deltaXML}), i.e.,
\begin{eqnarray}
\Delta X_{|\psi^{\mathrm{ML}}_\xi\rangle}=1.0998(\Delta X)^*_{min},
\end{eqnarray}
which shows an error less than $10\%$. Also, because of the
fuzziness of space, these maximal localization states are not
mutually orthogonal. It is worth to mention that, in this framework,
the expectation value of the kinetic energy operator $P^2/2m$ is
finite for both $|\psi_\lambda\rangle$ and
$|\psi^{\mathrm{ML}}_\xi\rangle$. Indeed we have
\begin{eqnarray}\label{posi22}
\left\langle\psi_\lambda\Big|\frac{P^2}{2m}\Big|\psi_\lambda\right\rangle=\frac{1}{10m\beta
},
\end{eqnarray}
and
\begin{eqnarray}
\left\langle\psi^{\mathrm{ML}}_\xi\Big|\frac{P^2}{2m}\Big|\psi^{\mathrm{ML}}_\xi\right\rangle=\frac{0.7345}{10m\beta
}.
\end{eqnarray}
These quantities for the KMM proposal are $\infty$ and $1/2m\beta$,
respectively.

\begin{figure}
\begin{center}
\includegraphics[width=7cm]{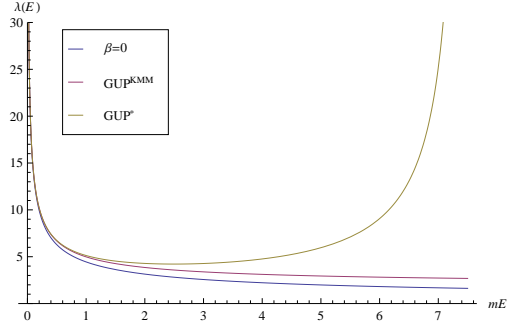}
\caption{\label{fig6}The wavelength of the quasiposition wave
function of a momentum eigenstate in ordinary quantum mechanics,
KMM's GUP and GUP* for $\beta=0.2$. }
\end{center}
\end{figure}

To find the quasiposition wave function $\psi(\xi)$, we  define
\begin{eqnarray}
\psi(\xi)\equiv\langle\psi^{\mathrm{ML}}_\xi|\psi\rangle,
\end{eqnarray}
where in the limit $\beta\rightarrow0$ it goes to the ordinary
position wave function $\psi(\xi)=\langle\xi|\psi\rangle$. Now the
transformation of the wave function in the momentum representation
into its counterpart quasiposition wave function is
\begin{eqnarray}
\psi(\xi)&=&\mathcal{N}\int_{-1/\sqrt{\beta}}^{+1/\sqrt{\beta}}\mathrm{d}p\,(1-\beta
p^2)\exp\bigg[\frac{i}{\hbar}\xi\left(p-\frac{\beta}{3}p^3\right)-\beta
\left(p^2-\frac{\beta}{2}p^4\right)\bigg]\psi(p).\label{psiF}
\end{eqnarray}
This relation shows that similar to the ordinary quantum mechanics
and the KMM proposal, the quasiposition wave function of a momentum
eigenstate $\psi_{\tilde{p}}(p)=\delta(p-\tilde{p})$ with energy
$E=\tilde{p}^2/2m$ is still a plane wave but with a modified
dispersion relation
\begin{eqnarray}\label{lambda}
\lambda(E)=\frac{2\pi\hbar}{\sqrt{2mE}\left(1-\frac{2}{3}m\beta
E\right)}=\frac{\lambda_{\mathrm{ord}}(E)}{1-\frac{2}{3}m\beta
E},
\end{eqnarray}
where $\lambda_{\mathrm{ord}}(E)=2\pi\hbar/\sqrt{2mE}$ is the
wavelength in the absence of GUP. In Fig.~\ref{fig6} we have
depicted $\lambda$ versus $mE$ in various scenarios. Since
Eq.~(\ref{lambda}) is bounded from below, there exists a nonzero
minimal wavelength. So the wavelength components smaller than
\begin{eqnarray}
\lambda_0=3\pi\hbar\sqrt{\beta}=\frac{3}{4}\pi\lambda_0^{\mathrm{KMM}},
\end{eqnarray}
are absent in the Fourier decomposition of the quasiposition wave
function of the physical states. Therefore, the maximal energy of a
momentum eigenstate is
\begin{eqnarray}\label{energymaxim}
E_{max}=\frac{3}{2m\beta}.
\end{eqnarray}

Since the transformation (\ref{psiF}) as the generalized Fourier
transformation is invertible, the transformation of a quasiposition
wave function into a momentum space wave function is given by
\begin{eqnarray}
\psi(p)&=&\frac{\mathcal{N}^{-1}}{2\pi\hbar}\int_{-\infty}^{+\infty}\mathrm{d}\xi\,\exp\bigg[\beta
\left(p^2-\frac{\beta}{2}p^4\right)\bigg]\exp\bigg[-\frac{i}{\hbar}\xi\left(p-\frac{\beta}{3}p^3\right)\bigg]
\psi(\xi).
\end{eqnarray}
Now the scalar product of states in terms of the quasiposition wave
functions reads
\begin{eqnarray}
\langle\phi|\psi\rangle&=&\int_{-1/\sqrt{\beta}}^{+1/\sqrt{\beta}}\mathrm{d}p\,(1-\beta
p^2)\phi^*(p)\psi(p),\nonumber\\
&=&\left(\frac{\mathcal{N}^{-1}}{2\pi\hbar}\right)^2\int_{-1/\sqrt{\beta}}^{+1/\sqrt{\beta}}\int_{-\infty}^{+\infty}\int_{-\infty}^{+\infty}\mathrm{d}p\,
\mathrm{d}\xi\,\mathrm{d}\xi'(1-\beta
p^2)\exp\bigg[2\beta\left(p^2-\frac{\beta}{2}p^4\right)\bigg]\nonumber\\
&&\times\exp\bigg[-\frac{i}{\hbar}(\xi-\xi')\left(p-\frac{\beta}{3}p^3\right)\bigg]
\phi^*(\xi)\psi(\xi').
\end{eqnarray}

\section{Harmonic oscillator}
In this section, we apply the developed formalism to the case of a
linear harmonic oscillator. Using the expression for the Hamiltonian
\begin{eqnarray}
H=\frac{P^2}{2m}+\frac{1}{2}m\omega^2X^2,
\end{eqnarray}
and the representation for $X$ and $P$, we obtain the following form
for the stationary state Schr\"odinger equation:
\begin{eqnarray}
\frac{d^2\psi(p)}{dp^2}+\frac{2\beta p}{1-\beta
p^2}\frac{d\psi(p)}{dp}+\left(1-\beta
p^2\right)^2\left(\epsilon-\eta^2p^2\right)\psi(p)=0,\label{SHOQM}
\end{eqnarray}
where $-1/\sqrt{\beta}\leq p\leq1/\sqrt{\beta}$ and
\begin{eqnarray}
\epsilon=\frac{2E}{m\hbar^2\omega^2},\hspace{1cm}\eta=\frac{1}{m\hbar\omega}.
\end{eqnarray}

\subsection{The quantum mechanical solution}
\begin{figure}
\begin{center}
\includegraphics[width=5cm]{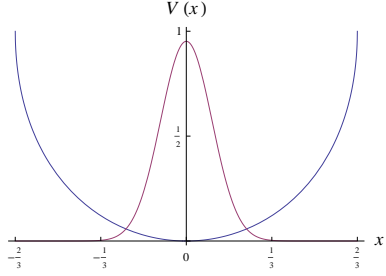}
\caption{\label{fig7}The effective potential $V(x)$ (blue line) and
the ground state wave function (red line) for $\eta'=100$. }
\end{center}
\end{figure}

Using the dimensionless variable $u=\sqrt{\beta}p$,
Eq.~(\ref{SHOQM}) can be written as
\begin{eqnarray}
\hspace{-1.5cm}\frac{d^2\psi(u)}{du^2}+\frac{2u}{1-u^2}\frac{d\psi(u)}{du}+\left(1-
u^2\right)^2\left(\epsilon'-\eta'^2u^2\right)\psi(u)=0,
\end{eqnarray}
where $-1\leq u\leq1$ and
\begin{eqnarray}
\epsilon'=\frac{\epsilon}{\beta},\hspace{1cm}\eta'=\frac{\eta}{\beta}.
\end{eqnarray}
Now by changing the variable to $x=u-(1/3)u^3$ we have
\begin{eqnarray}\label{schrodinger}
-\frac{d^2\psi(x)}{dx^2}+\eta'^2\,V(x)\psi(x)=\epsilon'\psi(x),
\end{eqnarray}
where $-2/3\leq x\leq2/3$ and
\begin{eqnarray}
\hspace{-1cm}V(x)=\left[\frac{1-i\sqrt{3}+(-2)^{1/3}\left(3x+\sqrt{9x^2-4}\right)^{2/3}}
{2^{2/3}\left(3x+\sqrt{9x^2-4}\right)^{1/3}}\right]^2,
\end{eqnarray}
is the effective potential which is real in this domain (see
Fig.~\ref{fig7}). The boundary condition now reads
\begin{eqnarray}\label{boundarycondition}
\psi(x)\Big|_{\pm\frac{2}{3}}=0.
\end{eqnarray}

To solve Eq.~(\ref{schrodinger}), we can expand the wave function in
terms of the particle in a box eigenfunctions. Since the potential
term $V(x)$ is an even function of $x$, to avoid large matrices, we
use
\begin{eqnarray}
\phi_m^{\mathrm{e}}(x)=\sqrt{\frac{1}{L}}\cos\left[\left(m-\frac{1}{2}\right)\frac{
\pi x}{L}\right],
\end{eqnarray}
and
\begin{eqnarray}
\phi_m^{\mathrm{o}}(x)=\sqrt{\frac{1}{L}}\sin\left(\frac{m \pi
x}{L}\right),
\end{eqnarray}
basis functions ($m=1,2,\ldots$) for even and odd
parity solutions, respectively, and write the wave function as
$\psi(x)= \sum_{m} A_{m} \phi_m(x)$ which vanishes at $\pm L$. Now
the boundary condition (\ref{boundarycondition}) reads $L=2/3$.

The approximate solutions are the eigenvalues and the eigenfunctions
of the $(N\times N)$ Hamiltonian matrix $\mathbf{H}_N$ in the form
\begin{equation}\label{even}
H_{mn}=\displaystyle\left(m-\frac{1}{2}\right)^2\frac{\pi^2}{L^2}\delta_{mn}
\nonumber +D_{mn}^{\mathrm{even}},
\end{equation}
and
\begin{equation}\label{odd}
H_{mn}=\displaystyle\frac{m^2\pi^2}{L^2}\delta_{mn}
+D_{mn}^{\mathrm{odd}},
\end{equation}
for even and odd states, respectively. Here, $\delta_{mn}$ is the
kronecker's delta and
\begin{eqnarray}
D_{mn}^{\mathrm{even}}=\frac{\eta'^2}{L}&&\int_{-L}^{L}V(x)\cos\left[\left(m-\frac{1}{2}\right)\frac{
\pi x}{L}\right]\cos\left[\left(n-\frac{1}{2}\right)\frac{ \pi
x}{L}\right]\mathrm{d}x,
\end{eqnarray}
\begin{eqnarray}
D_{mn}^{\mathrm{odd}}=\frac{\eta'^2}{L}\int_{-L}^{L}V(x)\sin\left(\frac{m
\pi x}{L}\right)\sin\left(\frac{n \pi x}{L}\right)\mathrm{d}x,
\end{eqnarray}
where $m$ and $n$ run from $1$ to $N$. In the usual diagonalization
scheme with the particle in box basis functions, we need to adjust
the domain $L$ with respect to the number of basis functions in such
way that the total error to be minimized \cite{pedramMolPhys}.
However, for our case, since the boundary condition
(\ref{boundarycondition}) has fixed the domain, i.e., $L=2/3$, the
accuracy of the solutions grows as the number of the basis
increases. In Table \ref{tab2} we have reported the first ten energy
eigenvalues of the harmonic oscillator in the GUP* framework. Indeed
$N=30$ basis functions suffices to obtain nearly accurate results
for the low lying energy eigenstates.

\begin{table}
\centering
\begin{tabular}{ccccc}\hline
$\hspace{.5cm}n\hspace{.5cm}$ & $\mathcal{E}^{\beta=0}_n$ & $\mathcal{E}^{(SC)}_n$    &$\mathcal{E}_n$  & $\frac{|\mathcal{E}_n-\mathcal{E}^{(SC)}_n|}{\mathcal{E}_n}$ \\\hline
0   &         1               &       1.00251               &   1.00509    &   $2.6\times10^{-3}$     \\
1   &         3               &       3.02284               &   3.02559    &   $9.1\times10^{-4}$     \\
2   &         5               &       5.06411               &   5.06704    &   $5.8\times10^{-4}$     \\
3   &         7               &       7.12698               &   7.13011    &   $4.4\times10^{-4}$     \\
4   &         9               &       9.21216               &   9.21550    &   $3.6\times10^{-4}$     \\
5   &         11              &       11.3204               &   11.3240    &   $3.2\times10^{-4}$     \\
6   &         13              &       13.4524               &   13.4563    &   $2.9\times10^{-4}$     \\
7   &         15              &       15.6091               &   15.6133    &   $2.7\times10^{-4}$     \\
8   &         17              &       17.7913               &   17.7958    &   $2.5\times10^{-4}$     \\
9   &         19              &       20.0000               &   20.0049    &   $2.4\times10^{-4}$
\\\hline
\end{tabular}
\caption{\label{tab2}The energy eigenvalues of the harmonic
oscillator in the GUP* framework. Here
$\mathcal{E}_n=\epsilon_n'/\eta'=\epsilon_n/\eta=2E_n/\hbar\omega$,
$N=30$, and $\eta'=100$.}
\end{table}

\subsection{The semiclassical solution}
\begin{figure}
\begin{center}
\includegraphics[width=7cm]{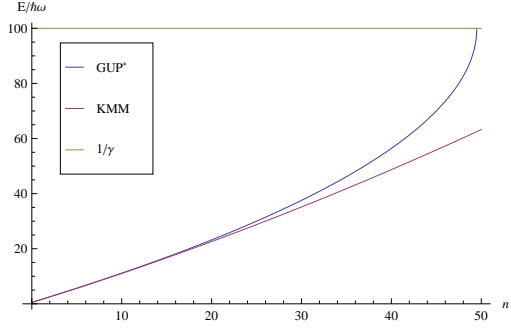}
\caption{\label{fig2}Comparing semiclassical harmonic oscillator
energy spectrum for KMM GUP (red line) and GUP* (blue line). We set
$\gamma=0.01$.}
\end{center}
\end{figure}

The total energy in terms of ordinary variables is
\begin{eqnarray}
E=\frac{p^2}{2m}+\frac{m\omega^2x^2}{2\left(1-\beta p^2\right)^2}.
\end{eqnarray}
To find the approximate energy eigenvalues of the above Hamiltonian,
we use the Wilson-Sommerfeld quantization rule in the form
\begin{eqnarray}
\oint
x\,\mathrm{d}p=\left(n+\frac{1}{2}\right)h,\hspace{1cm}n=0,1,2,\ldots,
\end{eqnarray}
where we have used $\oint \mathrm{d}(xp)=0=\oint
x\,\mathrm{d}p+\oint p\,\mathrm{d}x$. This integral can be written
as
\begin{eqnarray}
\oint x\,\mathrm{d}p=\frac{2}{m\omega}\int^{z}_{-z}\left(1-\beta
p^2\right)\sqrt{z^2-p^2}\,\mathrm{d}p,
\end{eqnarray}
where $z=\sqrt{2mE}$. So the semiclassical energy spectrum is given
by
\begin{eqnarray}
E_n^{(SC)}&=&\frac{1-\sqrt{1-2m\beta\hbar\omega\left(n+\frac{1}{2}\right)}}{m\beta},\label{Energy}\\
&=&-\frac{1}{8}\gamma\hbar\omega+\hbar\omega\left(n+\frac{1}{2}\right)\left(1+\frac{\gamma}{2}\right)+\frac{1}{2}\gamma
\hbar\omega n^2+\frac{1}{2}\gamma^2
\hbar\omega\left(n+\frac{1}{2}\right)^3+{\cal O}(\gamma^3),
\end{eqnarray}
where $\gamma=\beta m\hbar\omega=\eta'^{-1}$. As it is shown in the
appendix, the first three terms are similar to the energy spectrum
of the harmonic oscillator in the KMM framework. In Fig.~\ref{fig2},
we have depicted the energy spectrum in both GUP$^{\mathrm{KMM}}$
and GUP* frameworks. Note that, in GUP* scenario, the energy is also
bounded from above. Indeed, the maximum possible energy for the
harmonic oscillator is
\begin{eqnarray}
E_{max}^{(SC)}=\frac{1}{m\beta},
\end{eqnarray}
and the number of states ($N=n+1$) is finite, namely
\begin{eqnarray}
n_{max}=\left\lfloor\frac{1}{2\gamma}-\frac{1}{2}\right\rfloor,
\end{eqnarray}
where $\lfloor x\rfloor$ denotes the largest integer not greater
than $x$. So, to have at least one state, we should have
\begin{eqnarray}
\gamma\leq1.
\end{eqnarray}
As Fig.~\ref{fig2} shows, we have 50 states for $\gamma=0.01$.  The
first ten semiclassical energy eigenvalues are presented in Table
\ref{tab2}.\footnote{We used the relation
$\mathcal{E}^{(SC)}_n=2\eta'\Big(1-\sqrt{1-2\eta'^{-1}(n+1/2)}\Big)$.}
As the table shows the semiclassical results agree well with the
quantum mechanical energy spectrum. In fact, the relative error is
less than $3\times 10^{-3}$ even for the ground state. It is worth
to mention that a model with analogous properties has been studied
in the context of the nonrelativistic Snyder model in curved space
\cite{Mignemi}. Moreover, in the KMM framework, the bound states of
the relativistic particle in a box problem is also finite
\cite{pedramPLB}.

\subsection{The classical solution}
In the classical domain, the equations of motion are
\begin{eqnarray}
\dot{X}&=&\{X,H\}=\frac{P}{m\left(1-\beta P^2\right)}, \\
\dot{P}&=&\{P,H\}=-\frac{m\omega^2 X}{1-\beta P^2}.
\end{eqnarray}
The solutions to these equations are
\begin{eqnarray}
\omega
t&=&\left(1-\frac{\epsilon}{2}\right)\arccos\left(\frac{P(t)}{P_{max}}\right)-\frac{\beta}{2}P(t)\sqrt{P_{max}^2-P^2(t)}, \\
X(t)&=&-\frac{1-\beta
P^2(t)}{m\omega^2}\frac{\mathrm{d}P(t)}{\mathrm{d}t},
\end{eqnarray}
where
\begin{eqnarray}
\epsilon=2m\beta E,\hspace{1cm}P_{max}=\sqrt{2mE}.
\end{eqnarray}
To first-order in $\beta$ we have
\begin{eqnarray}
\hspace{-1cm}P(t)&=&P_{max}\left\{\cos\left[\left(1+\frac{\epsilon}{2}\right)\omega t\right]-\frac{\epsilon}{2}\sin^2\omega t\cos \omega t\right\}, \\
\hspace{-1cm}X(t)&=&X_{max}\left\{\left(1+\frac{\epsilon}{2}\right)\cos\left[\left(1+\frac{\epsilon}{2}\right)\omega
t\right]-\frac{\epsilon}{2}\sin^3\omega t\right\},
\end{eqnarray}
where $X_{max}=\displaystyle\sqrt{2E/m\omega^2}$. As we have
expected these results agree with the KMM proposal to
$\mathcal{O}(\beta)$ \cite{prd}.

It is straightforward to show that the infinitesimal phase space
volume between equal energy contours $E$ and $E+\mathrm{d}E$, and
equal time contours $t$ and $t+\mathrm{d}t$ can be written as
\begin{equation}
\mathrm{d}E\,\mathrm{d}t =(1-\beta P^2)\mathrm{d}X\,\mathrm{d}P.
\end{equation}
Now, since by definition the left hand side of this equation is time
independent, the right hand is also time independent.

\section{Conclusions}
In this Letter, we have presented a higher order generalized
uncertainty principle that implies both a minimal length uncertainty
and a maximal momentum proportional to $\hbar\sqrt{\beta}$ and
$1/\sqrt{\beta}$, respectively. We found the exact eigenfunctions of
the position operator and the quantum mechanical and semiclassical
energy spectrum of the harmonic oscillator and showed that the
energy spectrum is also bounded from above. Here we implemented a
momentum cutoff not through terms like $P^2$ on the right hand side
of the commutation relations. Instead, we implemented the momentum
cutoff through a function of $P$ with a singularity. So the momentum
space is cut into several sectors that decouple from each other. The
sectors are separated from each other at the singularities of the
function of $P$ that is used. Technically, we have inequivalent
irreducible representations of the commutation relations, one each
in each sector \cite{Kempf0}. This type of issue with the various
sectors can be avoided, as it is indicated in Ref.~\cite{Kempf2}.
Applied to our case, the trick would be to write the right hand side
of the commutation relation not as a fraction but instead to expand
it out as a geometric series. It has a finite radius of convergence
and that rules out all representations beyond the singularity. The
generalization of this GUP to $D$ dimensions which is
noncommutative, its invariant density of states, and its effects on
the blackbody radiation spectrum and the cosmological constant
problem are discussed in \cite{pedramF}.

\appendix

\section{Harmonic oscillator spectrum in the KMM framework}
In the context of the KMM proposal, the total energy in terms of
ordinary variables is given by
\begin{eqnarray}
E=\frac{p^2}{2m}+\frac{1}{2}m\omega^2\left(1+\beta p^2\right)^2x^2.
\end{eqnarray}
Now the Wilson-Sommerfeld integral can be written as
\begin{eqnarray}
\oint
x\,\mathrm{d}p=\frac{2}{m\omega}\int^{z}_{-z}\frac{\sqrt{z^2-p^2}}{1+\beta
p^2}\mathrm{d}p=\left(n+\frac{1}{2}\right)h,
\end{eqnarray}
where $z=\sqrt{2mE}$. So the semiclassical energy spectrum is given
by
\begin{eqnarray}\label{energyKMM}
\hspace{-1cm}E_n^{(SC)}=-\frac{1}{8}\gamma\hbar\omega+\hbar\omega\left(n+\frac{1}{2}\right)\left(1+\frac{\gamma}{2}\right)+\frac{1}{2}\gamma
\hbar\omega n^2,
\end{eqnarray}
where $\gamma=\beta m\hbar\omega$. This result agrees (up to a
constant) with the exact solution to first order of the GUP
parameter \cite{7}
\begin{eqnarray}
E_n^{exact}&=&\hbar\omega\left(n+\frac{1}{2}\right)\left(\sqrt{1+\gamma^2/4}+\gamma/2\right)+\frac{1}{2}\gamma \hbar\omega n^2\nonumber\\
&=&E_n^{(SC)}+\frac{1}{8}\gamma\hbar\omega+\mathcal{O}(\gamma^2),
\end{eqnarray}
and gives the correct $n^2$ dependence behavior. An alternative
derivation of Eq.~(\ref{energyKMM}) is also presented in
Ref.~\cite{pedramPRD}.

\section*{Acknowledgement}
I am very grateful to Achim Kempf, Rajesh R.~Parwani, and Kourosh
Nozari for fruitful discussions and suggestions.

\end{document}